\documentclass[12pt,a4paper]{article}
\usepackage{russian,english}
\usepackage{latexsym}
\usepackage{graphicx}

\title{\large\bf Electron-hole pairs multiquantum\\
  generation mechanizm in the process\\
  of excitation of $ZnSCdS-Cu$ by an atomic hydrogen}
\author{{\large Yu.A.Sivov, Yu.I.Tyurin, V.D.Choruzhii}\\
{\small\bf Tomsk Polytechnical University, Rossia}}
\date{}
\begin{document}
\maketitle
\begin{abstract}
The luminecsence of $ZnSCdS-Cu$ phosphors of various $CdS$
consentration ($0-40\%$) is under investigation. The
luminescence was exited by hydrogen atoms during adsorbtion of thermal energy during adsorbtion  and impact recombination-heterogeneous
chemiluminenscence($HCL$) are discussed. The monotone
lowering of the conductivity zone bottom level with respect
to the valence zone ceiling (from $3,7 eV$ to $3,2 eV$) with
increasing $CdS$ concentration (from $0 - 40\%$). $HCL$
intensity with respect to $CdS$ (from 3,7 eV to 3,2 eV)-
concentration ($E_g$)grows rapidly.
 The
dependence of $HCL$ intensity on the forbidden zone width can
be explained by the action of electron - hole pairs
multiquontum generation mechanism.
\end{abstract}

n1. The termal energy atoms and molecules trapped on surface
are capable to perform high-frecuency intermolecular and
adsorbtional oscillations. The adsorbtion potential depth of
chemical active particles on the uviol surfaces variates from
$q = 1 eV$ up to $10 eV$ (the frecuencies correspondingly -
from $\omega_0 = 10^{14}s^{-1}$ up to $10^{15}s^{-1}$.
 One-three photon energy scattering into a crystal
is a very effective process near the border of continious
energy spectrum. The relaxation rate $Ø\Gamma_{vph} = 10^8 -
10^{11} s^{-1}$ \cite{p1,p2}. Having released the energy
$\hbar\omega_0 >>kT$ during the time-period $(2 \div 5)
\Gamma_{vph}^{-1}$, a particle is caught by the surface, but
at the same time it has a large energy exess in its
adsorbtion binding ($q - \hbar\omega_0 \geq \hbar\omega_0$).
The following energy release into the crystalline lattice is
inhibited by the multiquantum processes (($E_V -
E_{vi})/(\hbar\omega_{ph})\geq 3  \div 5;
 \Gamma_{vph}= 10^2\div  10^7 s^{-1}; \hbar\omega_{ph}$-phonon
energy). In this case, the nonphonon relaxation processes of
energy release into an adsorbtion layer, to ionized surface
states, or to the electrons of impurity, Bloch and Tamm
states are effective \cite{p3,p4}.

The role of the crystal electron states in the energy
accomodation process taking place in the gas-surface
interaction is widely observed during luminescence of solids
located into the active gas medium (free atoms,
radicals).This kind of nonequilibrium luminescence is known
to be called heterogeneous chemiluminescence $HCl$ \cite{p5}.
Two modifications of $HCL$ are known: radicalo-recombination
and adsorbtion luminescence ( $RRL$ and $AL$) $\cite{p5}-\cite{p7}$.
$RRL$ is observed in the free atoms recombination process,
$AL$ - in the process of adsorption.


n2. This paper presents the experimental results of $RRL$ of
$ZnSCdS-Cu$ with the various cosentrations of $CdS (0 - 40\%
)$, in the atomic hydrodgen medium. The atomic hydrodgen is
produced by dissociation of the molecular hydrodgen during
high-frecuency nonelectrode electric charge. In its turn, the
molecular hydrodgen is produced by diffusion through a heated
palladium.

Kinetic, temperature and spectral characteristic of
luminencence are studied with the help of photoelectric
amplifier ($PEA - 84$) attached to the output of a
high-aperture monochromator, sometimes the interference
filters are used.

\begin{figure}
\includegraphics[width=6cm,height=3.5cm]{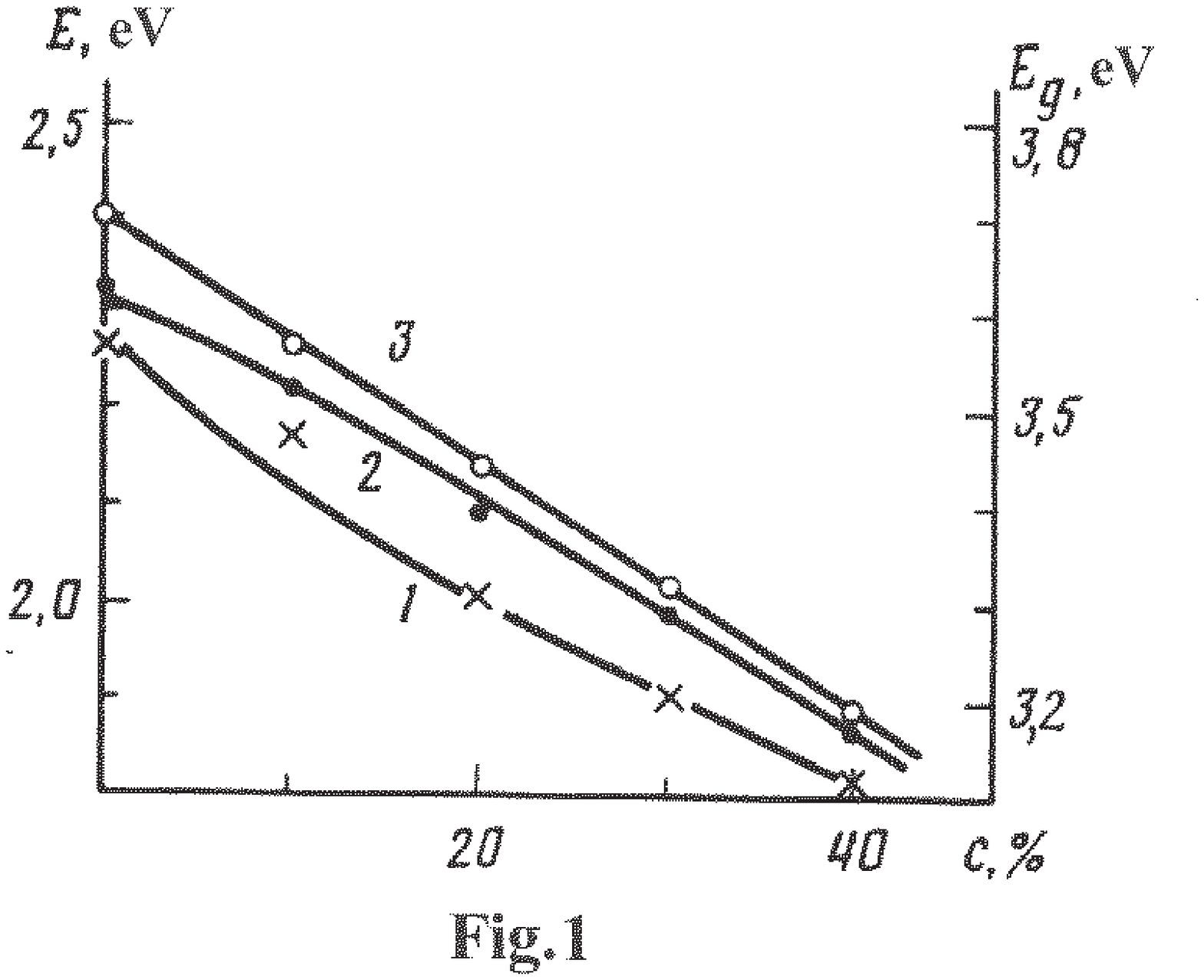}
\hfill
\includegraphics[width=6cm,height=3.5cm]{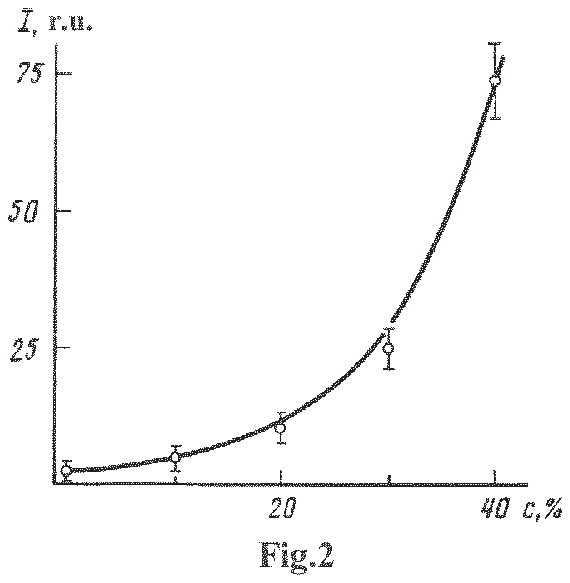}
\\
\parbox[t]{6cm}{\caption{Dependence of the pozition of luminiscence spectrum
maxima (1,2) and forbidden zone width (3) of $ZnSCdS-Cu$
phosphor on the $CdS$ concentration; 1- photoluminicence,
2-heterogeneous chemiluminescence in the atomic
hydrodgen.}\label{f1}} \hfill
\parbox[t]{6cm}{\caption{Dependence of the $HCL$ intensits of the $ZnSCdS-Cu$
phosphor on the $CdS$ concentration. Points - are the
experimental results, the line represents the theoretical
colculations.}\label{f2}}
\end{figure}

In Fig.1, position of spectrum maxima of photo and heterogenerous
chemiluminescence and  the widths of forbidden zone $E_g =
E_g (ZnS)[1 - C(CdS)] + E_g (CdS)C(CdS)$ of phosphor
$ZnSCdS-Cu$ are given as functions of concentration of
$CdS(C(CdS) = 0 - 40\%$). If can be noted that the rate of
spectrum maxima and $E_g(C)$ displayments are practically thesame. $HCL$ intensity with respect to $CdS$ - consentration($E_g$ decreases) grows rapidly (Fig.2).

For the $ZnSCdS-Cu$ phosphor containing the variable
quantities of $CdS(0-40\%$) it is established  the
$Cu^{2+}$-centre radiation spectrum is displaced to long
wavelength (from $2,4 eV$ to $1,9 eV$). It is caused by
monotone lowering of the conductity zone bottom level with
respect to the valence zone ceiling (from $3,7 eV$ to $3,2
eV$), pozition of the $Cu^{2+}$ - centre with respect to the
valence zone ceiling being the same (theory of indirect
activation \cite{p5}).

The dependence of the $RRL$ intensity of the $ZnSCdS-Cu$
phospor on the $CdS$ concentration can be explained by the
action of electron-hole pairs multiquontum generation
mechanism.


n3. Because of the oscillation unharmonism and dependence of
the dipole (quadrupole) momentum from binding on the internuclear
coordinates, the interaction between the oscillotory excited
dipole and the crystall elerctron states leads to
transmiision of several oscillatory quanta to electrons states at
the same time.

If the Morze potential is chosen to describe the unharmonic
oscillator, then the rate of the multiquantum oscillatory
electron transitions by the zone-zone excitations can be
expressed \cite{p4} as follows
\begin{equation}
\label{e1} \Gamma_{ev} = \displaystyle\frac{4m_2\theta^2
f(m_r \hbar \omega_0)^{1/2}}{\hbar \omega_0 M m_e E_g}
[\displaystyle\frac{\mu (r_0)e(\varepsilon
+2)}{3a^{3/2}\varepsilon}]^2
\psi(q/E_g)exp(-\displaystyle\frac{E_g}{\hbar \omega_0}P).
\end{equation}
Here: $m_e$ - mass of electron, e - elementary electrical
charge, $\omega_0 = \alpha (2q/m)^{1/2}$ - oscillator's
cyclic frequency, $\mu$ - dipole moment of binding, f -
oscillatory force, $\epsilon$ - permittivity, corresponding
to the transmission frequency, q - adsorbtion potential
depth, $E_g$ - width of forbidden zone, $m_r$ - equivalent
effective mass of an electron - hole pair, a -minimal
distance of the energy transportation.
\begin{equation} \label{e2} \psi(x) =
x^2(1-x^{-1/2})(1 + (2x)^{-1/2})
\end{equation}
\begin{equation}
 \label{e3}
 p = z\ln{z/(z-1)},
z = (4q + \hbar \omega_0)/(E_g + \hbar \omega_0).
\end{equation}
$(\ref{e1})$ describes the oscillatory electron transition
(inside the static region of a dipole) produced by the dipole
interaction of a valence zone electron with the
electromagnetic field of an unharmonic oscillator. To
describe the dipole - quadrupole interaction, it is necessary
to change the expression inside the rectangle parenthesises
by $[eD(r_0)/5a^{5/2}]^2$.

The $(\ref{e1})$ and $(\ref{e3})$ describing the exitation
rate can be used to evaluate the dependence of the phosphor
$ZnSCdS-Cu$ luminiscence intensity on the $CdS$ concentration
(the width of the phosphor`s forbidden zone), (Fig.2).
Intensity of $RRL_H$ depends on the exitation cross - section
\cite{p9} and is proportional to the rate of the
nonequilibrium electron state generation:
\begin{equation}
\label{e4}
 I \sim \displaystyle\frac{\Gamma_{ev}}{(\Gamma_{ev} +
\Gamma_{v})(1 + \tau \omega}
 (\jmath\sigma_2 N_1 B_{RRL} +
\jmath\sigma_1NB_{Al}.
\end{equation}
Here: $\sigma_1$-adsorbtion cross-section, $\sigma_2$ -
recombination cross - section, $N, N_1$ - correspondingly
concentration of $AL$ and $RPL$ exitation centres,
$B_{AL},B_{RRL}$ - quantum yield of luminescence outputs of $AL$ and
$RRL$ exitation centres, $(1 + \tau \omega)^{-1}$ -
concentration extinguish factor, $\Gamma_v + \Gamma_{ev}$ -
total relaxation rate, $\jmath$ - atom flux density.

These quantities does not (or in a rather small degree)
depend on the energy width $E_g$.

For this situations according $(\ref{e1}-\ref{e4})$ we get:
 \begin{equation}
 \label{e5}
 I_c = a/E_g \Psi (q/E_g)exp(\displaystyle\frac{-E_g}{\hbar\omega_0}P)
 \end{equation}
 \begin{equation}
 \label{e6}
 E_g = [3,6 - 1,1 C(CdS)] eV
 \end{equation}
where $C(CdS)$ concentration $CdS$ inside the $ZnSCdS-Cu$
phosphor.

The theoretical curve $I(C)$ coincides (in the limits of
experimental errors) with the experimental results (see
$Fig.2). (\hbar\omega_0 = 0,32 eV, q = 4 eV$). The increment
of $I(C)$ is not exponential and depends strongly on the
factor ($1/E_g)\Psi(q/E_g)$.

When the $CdS$ concentration is great ($C > 60\%$), the $HCL$
output and intensity tend to be lower. It is caused by
increasing of the consentration extinguish (factor $(1 +
\tau\omega)^{-1}$) and rate of the low - energy relaxation
$\Gamma_v$ of the oscillatory exitation binding $(H -
Me^{2+})^v - SU, (H - H)^v - SU.$


\begin{thebibliography}{000}
\addcontentsline{toc}{chapter}{Referenses}
\bibitem{p1}
M.A.Koshushner. //Theoretical problems of chemical physic.
M.:"Science", 1982. P.283.
\bibitem{p2}
V.P.Zhdanov V.P.,Zamaraev K.I. //Catal.Sci.Eng., 1982. V.24.
N3. P.273.
\bibitem{p3}
B.R.Shub. Heterogeneous relaxation of reactions chemical
internal energy molecules and heterogeneous processes at the
solid state surface. Doct. diss. chem. science. M. :Inst.of
chem.phys., 1983.
\bibitem{p4}
Tyurin Yu.I. //Surface, 1986. N9. P.115.
\bibitem{p5}
V.V.Styrov. Energy reactions heterogeneous chemical
relaxation with participation of crystall system electrons.
Preprint 19-78. Novosibirsk. Inst.of semiconductor physics SO
AN SSSR, 1978.
\bibitem{p6}
V.A.Sokolov, A.N.Gorban. //Luminescence and adsorbtion. M.
:"Science", 1969. P.188.
\bibitem{p7}
Yu.N.Ruphov. //Problems of kinetic and catalisys.
M.:"Science", 1978. V.17. P.63.
\bibitem{p8}
A.M.Gurvisch. Introduction at chemistry physic of
crystallophophors. M.:"High.sch.", 1982. P.376.
\bibitem{p9}
Yu.I.Tyurin Yu.A., V.V.Styrov V.V. //Chem.phys. 1984. V.3.
V.1. P.85.

\end{thebibliography}
\end{document}